# Macro-trends in research on the central dogma of molecular biology

Sepehr Ehsani

Department of Laboratory Medicine and Pathobiology, and Tanz Centre for Research in Neurodegenerative Diseases, University of Toronto, Toronto, Ontario M5S 3H2, Canada

*Present address:* Whitehead Institute for Biomedical Research, and MIT Computer Science and Artificial Intelligence Laboratory, Cambridge, Massachusetts 02142, United States

ehsani@csail.mit.edu

**ABSTRACT**

*The central dogma of molecular biology, formulated more than five decades ago, compartmentalized information exchange in the cell into the DNA, RNA and protein domains. This formalization has served as an implicit thematic distinguisher for cell biological research ever since. However, a clear account of the distribution of research across this formalization over time does not exist. Abstracts of >3.5 million publications focusing on the cell from 1975 to 2011 were analyzed for the frequency of 100 single-word DNA-, RNA- and protein-centric search terms and amalgamated to produce domain- and subdomain-specific trends. A preponderance of protein- over DNA- and in turn over RNA-centric terms as a percentage of the total word count is evident until the early 1990s, at which point the trends for protein and DNA begin to coalesce while RNA percentages remain relatively unchanged. This term-based census provides a yearly snapshot of the distribution of research interests across the three domains of the central dogma of molecular biology. A frequency chart of the most dominantly-studied elements of the periodic table is provided as an addendum.*

**MAIN TEXT**

The central dogma of molecular biology, consisting of the tripartite alignment of DNA, RNA and protein, was formalized in 1970 [1] (although first put forth in 1958) and has encapsulated the main domains of focus in biological research. Although the boundaries between these domains have become ever more intangible with discoveries such as DNA-protein complexes in the form of nucleosomes [2] or RNA-protein entities such as the ribosome [3], a given research project in the molecular biology laboratory can still be categorized as mainly emphasizing or at least partially involving one of the components of the central dogma more than the remaining two constituents.

It would be informative to determine the extent to which the field of molecular biology has emphasized DNA, RNA and protein as its focus of attention and how the emphasis might have changed over time in the form of a census. Text-mining could be a suitable approach to analyze such patterns in the published literature [4,5]. To that end, 3,767,522 abstracts of all articles containing the word “cell” or “cells” in any field or as a Medical Subject Heading (MeSH) were extracted from the Medical Literature Analysis and Retrieval System Online (MEDLINE) database for a 37-year period from 1975 to 2011, inclusive. Publications indexed prior to 1975 do not necessarily reflect the expected trajectory of increasing annual publication numbers and were not included for the purpose of this study. An in-house word-frequency Java program was used to determine the annual occurrence of a total of 100 DNA-, RNA- and protein-centric terms and their plural/adjectival derivatives, in addition to the total yearly word counts (**Table S1**). Although some of the selected search terms will inevitably capture minor aspects of a domain other than the one assigned, such crossovers will be negligible in the scale and theme of the current search. Moreover, to ensure consistency across different abstracts, the search terms were single words only, and referred to biological entities (e.g., genome) rather than techniques, processes or phenomena (e.g., sequencing, transcription, etc.).

The total number of words, after adjustments for the abstract number and PubMed ID line, was 939,309,404, equalling approximately 250 words per abstract across the 37-year period (**Fig. 1**). The slope of the total word count, however, is greater than that of the total publications per year, indicating a general trend towards longer abstracts. Whereas the average abstract length from 1975 to 1979 is 180, it increases to 271 words between 2007 and 2011.

The cumulative percentage of protein-centric terms relative to the total annual word count in abstracts from 1975 is 0.52%, whereas it is 0.32% for DNA-centric and 0.13% for RNA-centric terms (**Fig. 2A**). Protein and DNA terms show an increase until the early 1990s, followed by a decline in the late 1990s for protein and mid-2000s for DNA. The trend for RNA shows a relatively slighter variation in the 37-year period. In 2011, the relative percentages for protein, DNA and RNA had changed to 0.47%, 0.46% and 0.16%, respectively. It should be noted that although the number of publications in each domain has increased substantially every year, the percentage of search terms per total words indicates the relative focus of the field of molecular biology on each part of the central dogma. Overall, it is evident that the protein component of the central dogma of molecular biology received greater attention in research in the 1970s, followed by DNA and RNA. With a gradual decline in protein-centric percentages and relatively constant increase in DNA terms, the protein and DNA domains appear to have garnered equal focus in the 2000s. The relative attention on the RNA domain seems to have remained less fluctuating.

It will be useful to determine if the trend observed for each domain is mainly carried by one or more subdomains. 'Gene' and related terms are the predominant trendsetters for DNA-centric words (**Fig. 2B**). Other than 'genome' and 'epigenetics' (and related terms), which show a slow rise from early-2000s onward, all other terms have a negative or near-zero slope. The trend for RNA-centric words seems to be mainly carried in phases by three different subcategories: from 1975 to the mid-1980s by 'RNA', from the mid-1980s to the early 2000s by 'mRNA' (messenger RNA), and from then on by 'siRNA' (small interfering RNA) and 'miRNA' (microRNA) (**Fig. 2C**). The main term in the protein-centric trend is, unsurprisingly, 'protein' and related terms (**Fig. 2D**). However, the key contributors to the shape of the trend in the 1970s and 1980s are 'enzyme' and 'antibody', respectively, both of which demonstrate a steady decline in subsequent decades. The 'proteome' subdomain shows a relatively small rise in the past decade.

All 100 search terms account for an average of 1.15% of the total words in the abstracts with a standard deviation of 0.11%. If the three constituents of the central dogma encompass the principal avenues of information flow in the cell, it might then be plausible to assume that the cumulative percentage of all DNA-, RNA- and protein-related terms should be constant on a year-by-year basis, even in cases of new discoveries within the dogma which spawn new research directions and publications. Although the current census appears to corroborate this assumption to an extent, the sums of all percentages span a minimum of 0.95% in 1977 and a maximum of 1.32% in 1995 (**Fig. S1**). It is possible to speculate that publications analyzed in this study have at different times relied on biological phenomena outside of the tripartite axis (e.g., phospholipids) to complement their explanation of a given aspect of the cell's information system.

To control for the accuracy of the search algorithm, the frequency of the articles 'a', 'an' and 'the' were calculated (**Fig. S2**). The indefinite articles show a consistent percentage throughout the search period, with 'a' and 'an' representing 1.5% and 0.3% of the total words, respectively. Interestingly, the usage of the definite article 'the' shows a negative linear trend, with a decrease from 6.0% in 1975 to 3.8% in 2011.

Finally, to determine the relative focus of the field of molecular and cellular biology on the natural elements, the frequencies of occurrence of the 118 elements in the periodic table were calculated (**Fig. S3**). Calcium, oxygen, lead, iron, sodium, carbon, zinc, hydrogen, gold, copper and potassium represented the most frequently mentioned elements at present, respectively (**Fig. S3A**). Relative to other elements, calcium and sodium demonstrate a decrease in occurrence from the late 1980s to 2011 (**Fig. S3B**).

## FIGURE LEGENDS

**Figure 1. Overview of analyzed abstracts.** The total number of publications (bar graph, left vertical axis) shows a steady rise from tens of thousands in the late 1970s to hundreds of thousands in the late 2000s. The total number of words for each year (line graph, right vertical axis), however, increases more rapidly over the same time period and indicates a trend toward longer abstract lengths.

**Figure 2. Central dogma domain trends from 1975 to 2011.** Protein-centric research appears to have dominated the field of molecular biology in the late 1970s, followed by DNA-centric and RNA-centric research (**A**). With a steady rise in DNA-centric terms and decline in protein-centric terms, the DNA and protein domains converge on equal percentages relative to total words in the 2000s. The breakdown for individual trends demonstrating the contribution of each subdomain appears in **B**, **C** and **D**.

**Figure S1. Annual sums of percentages.** A revised **Fig. 2A** includes a cumulative line graph of the yearly DNA-, RNA- and protein-centric percentages, which range from 0.95% (1977) to 1.32% (1995).

**Figure S2. Control graph of the frequency of definite/indefinite articles.** The indefinite articles ‘a’ and ‘an’ demonstrate a steady frequency over the period 1975-2011, whereas the definite article ‘the’ decreases from 6.0% to 3.8% over the same period.

**Figure S3. Frequency of the elements.** The absolute (**A**) and relative (**B**) occurrences of elements 1-118 were calculated. Two alternative spellings were considered for four elements: aluminium/aluminum, phosphorus/phosphorous, sulfur/sulphur and caesium/cesium. The elements francium, berkelium, einsteinium, mendelevium, nobelium, lawrencium, rutherfordium, dubnium, seaborgium, bohrium, hassium, meitnerium, darmstadtium, roentgenium, copernicium, ununtrium, flerovium, ununpentium, livermorium, ununseptium and ununoctium did not occur in any abstract.

**Figure 1**

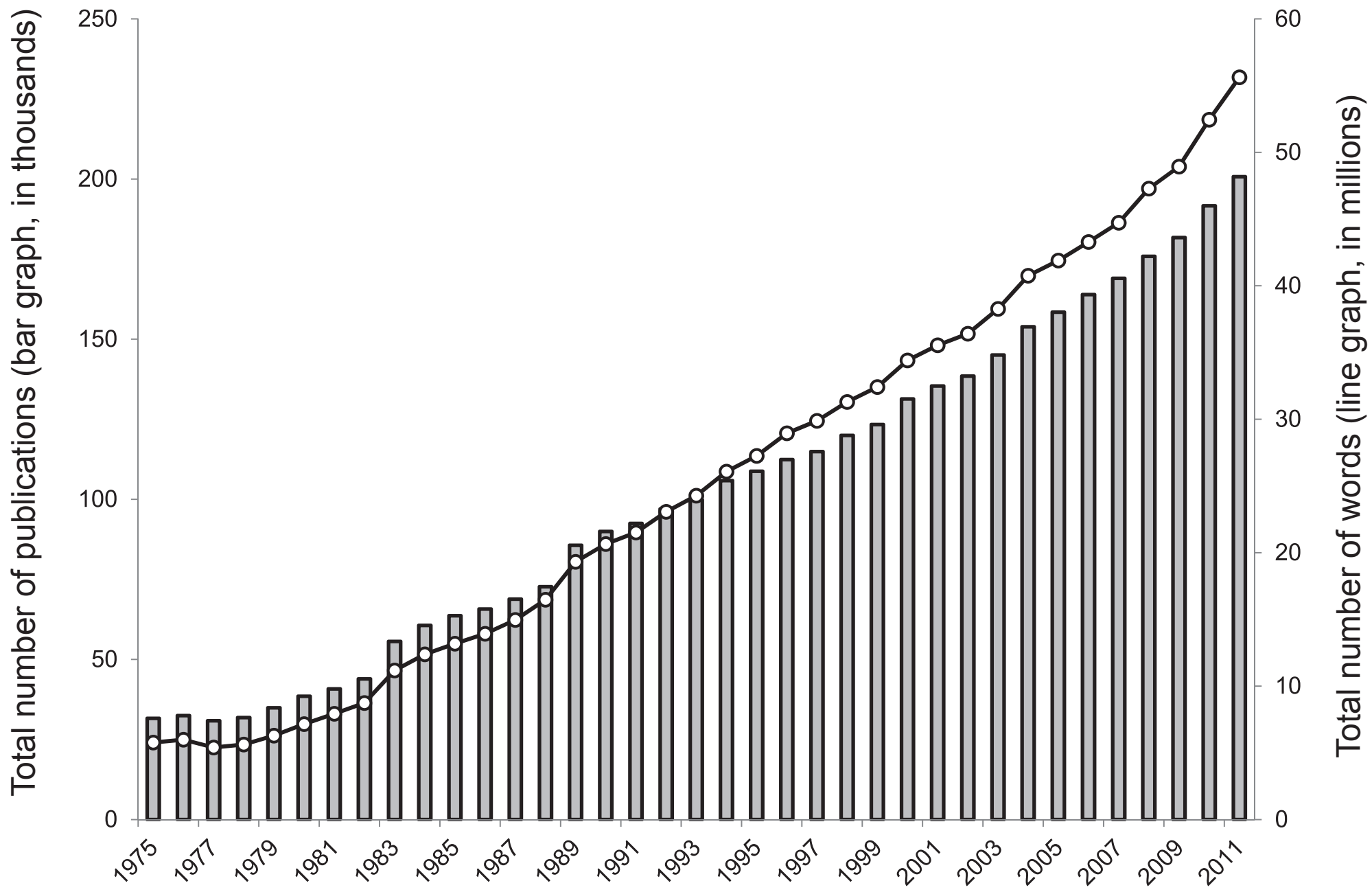

Total number of publications (bar graph, in thousands)
Total number of words (line graph, in millions)
250
200
150
100
50
0
60
50
40
30
20
10
0
1975
1977
1979
1981
1983
1985
1987
1989
1991
1993
1995
1997
1999
2001
2003
2005
2007
2009
2011

**Figure 2**

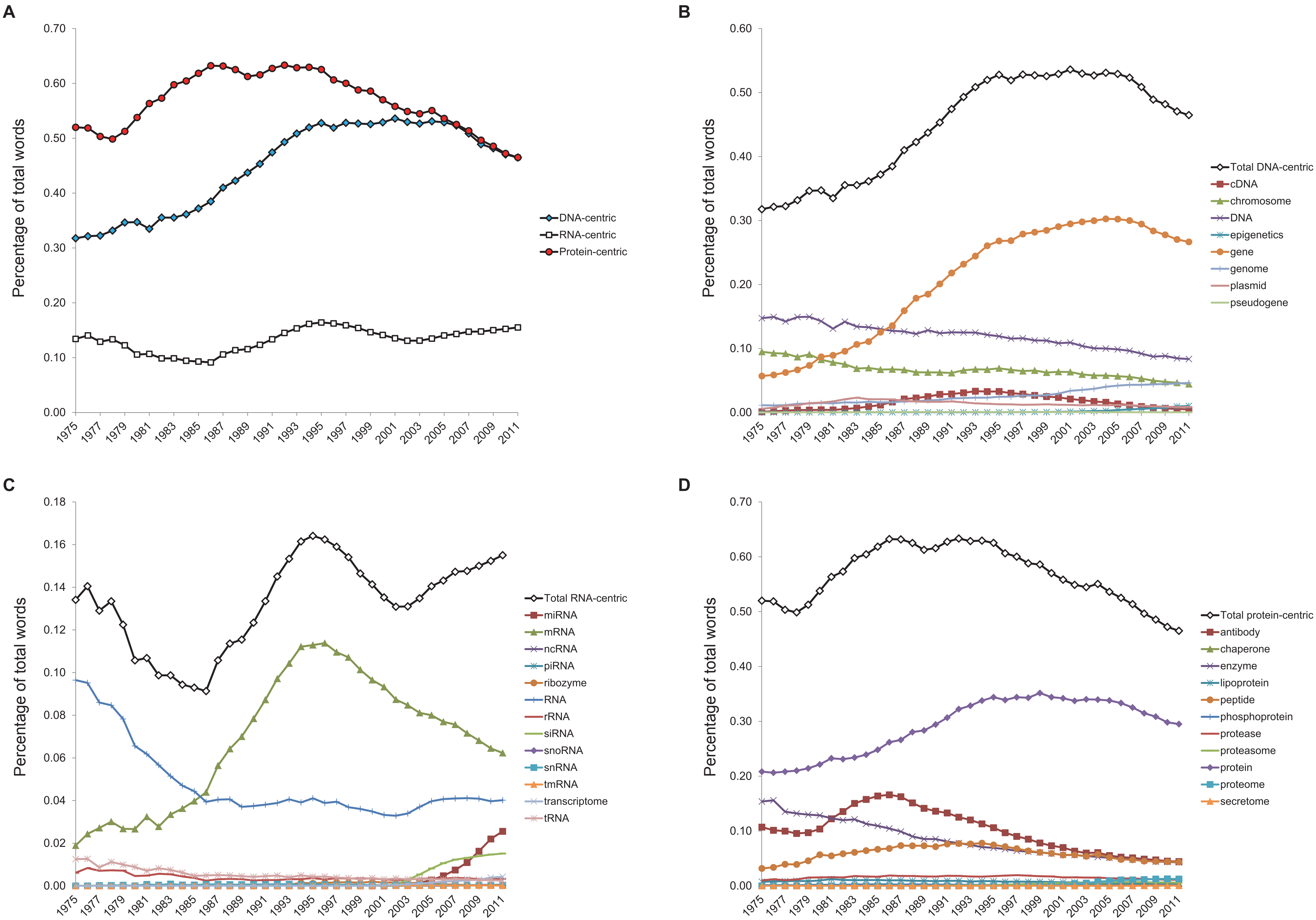
A
B
C
D
Percentage of total words
DNA-centric
RNA-centric
Protein-centric
Total DNA-centric
cDNA
chromosome
DNA
epigenetics
gene
genome
plasmid
pseudogene
Total RNA-centric
miRNA
mRNA
ncRNA
piRNA
ribozyme
RNA
rRNA
siRNA
snoRNA
snRNA
tmRNA
transcriptome
tRNA
Total protein-centric
antibody
chaperone
enzyme
lipoprotein
peptide
phosphoprotein
protease
proteasome
protein
proteome
secretome

**Table S1.** Annual publications, total word counts and DNA-, RNA- and protein-centric word counts

| | 1975 | 1976 | 1977 | 1978 | 1979 | 1980 | 1981 | 1982 | 1983 | 1984 | 1985 | 1986 | 1987 | 1988 | 1989 | 1990 | 1991 | 1992 | 1993 |
|---|---|---|---|---|---|---|---|---|---|---|---|---|---|---|---|---|---|---|---|
| **Publications ("cell")** | 31607 | 32446 | 30834 | 31857 | 34881 | 38488 | 40798 | 43919 | 55610 | 60644 | 63637 | 65755 | 68811 | 72707 | 85631 | 89973 | 92468 | 96988 | 99833 |
| **Total words** | 5774934 | 5990895 | 5404861 | 5629292 | 6288805 | 7159837 | 7930864 | 8735668 | 11170969 | 12383537 | 13184025 | 13926819 | 14958598 | 16478895 | 19317118 | 20651136 | 21514072 | 23060053 | 24273183 |
| cDNA | 53 | 117 | 157 | 209 | 231 | 284 | 301 | 446 | 711 | 1055 | 1433 | 2094 | 2868 | 3355 | 4237 | 5115 | 5449 | 6167 | 6938 |
| cDNAs | 11 | 3 | 7 | 10 | 7 | 21 | 14 | 31 | 63 | 86 | 116 | 188 | 287 | 386 | 549 | 737 | 765 | 920 | 1131 |
| chromatin | 1293 | 1535 | 1138 | 1176 | 1277 | 1373 | 1409 | 1386 | 1670 | 1806 | 1597 | 1489 | 1412 | 1337 | 1526 | 1419 | 1383 | 1458 | 1447 |
| chromatins | 26 | 46 | 21 | 18 | 22 | 19 | 9 | 13 | 23 | 14 | 11 | 11 | 8 | 13 | 4 | 5 | 4 | 11 | 11 |
| chromosomal | 635 | 688 | 616 | 590 | 757 | 830 | 787 | 1058 | 1142 | 1204 | 1468 | 1504 | 1686 | 1736 | 2098 | 2324 | 2286 | 2567 | 2772 |
| chromosome | 2149 | 1988 | 2025 | 2001 | 2361 | 2403 | 2681 | 2730 | 3387 | 4008 | 4046 | 4579 | 4990 | 5369 | 6407 | 6827 | 7222 | 8530 | 9357 |
| chromosomes | 1377 | 1298 | 1172 | 1096 | 1289 | 1285 | 1320 | 1403 | 1454 | 1581 | 1716 | 1815 | 1867 | 1906 | 2073 | 2349 | 2384 | 2646 | 2785 |
| DNA | 8418 | 8827 | 7609 | 8263 | 9245 | 10027 | 10175 | 12114 | 14697 | 16217 | 16873 | 17459 | 18611 | 19866 | 24443 | 25189 | 26528 | 28368 | 29920 |
| DNAs | 93 | 116 | 81 | 133 | 160 | 192 | 214 | 272 | 294 | 275 | 303 | 334 | 325 | 370 | 367 | 398 | 418 | 456 | 377 |
| epigenetic | 8 | 5 | 15 | 12 | 13 | 17 | 15 | 20 | 23 | 48 | 27 | 49 | 31 | 36 | 46 | 76 | 62 | 73 | 86 |
| epigenetics | 0 | 0 | 0 | 0 | 0 | 0 | 0 | 0 | 0 | 0 | 0 | 0 | 0 | 0 | 1 | 1 | 0 | 0 | 2 |
| epigenome | 0 | 0 | 0 | 0 | 0 | 0 | 0 | 0 | 0 | 0 | 0 | 0 | 0 | 0 | 0 | 0 | 0 | 0 | 0 |
| epigenomes | 0 | 0 | 0 | 0 | 0 | 0 | 0 | 0 | 0 | 0 | 0 | 0 | 0 | 0 | 0 | 0 | 0 | 0 | 0 |
| epigenomic | 1 | 0 | 0 | 0 | 0 | 0 | 0 | 0 | 0 | 0 | 1 | 0 | 2 | 0 | 0 | 0 | 0 | 0 | 1 |
| epigenomics | 0 | 0 | 0 | 0 | 0 | 0 | 0 | 0 | 0 | 0 | 0 | 0 | 0 | 0 | 0 | 0 | 0 | 0 | 0 |
| exome | 0 | 0 | 0 | 0 | 0 | 0 | 0 | 0 | 0 | 0 | 0 | 0 | 0 | 0 | 0 | 0 | 0 | 0 | 0 |
| exomes | 0 | 0 | 0 | 0 | 0 | 0 | 0 | 0 | 0 | 0 | 0 | 0 | 0 | 0 | 0 | 0 | 0 | 0 | 0 |
| exomic | 0 | 0 | 0 | 0 | 0 | 0 | 0 | 0 | 0 | 0 | 0 | 0 | 0 | 0 | 0 | 0 | 0 | 0 | 0 |
| exomics | 0 | 0 | 0 | 0 | 0 | 0 | 0 | 0 | 0 | 0 | 0 | 0 | 0 | 0 | 0 | 0 | 0 | 0 | 0 |
| gene | 1331 | 1422 | 1351 | 1599 | 2187 | 3004 | 3564 | 4390 | 6764 | 8188 | 10209 | 11854 | 14942 | 18072 | 22209 | 25992 | 29671 | 34121 | 38042 |
| genes | 707 | 864 | 861 | 957 | 1118 | 1655 | 1917 | 2246 | 3058 | 3474 | 4169 | 4759 | 5845 | 7094 | 8316 | 9798 | 11035 | 12269 | 13195 |
| genetic | 1103 | 1082 | 1031 | 1076 | 1174 | 1407 | 1434 | 1586 | 1877 | 1896 | 2004 | 2068 | 2246 | 2560 | 3239 | 3497 | 3819 | 4357 | 5186 |
| genetics | 152 | 149 | 137 | 120 | 146 | 162 | 169 | 134 | 185 | 169 | 161 | 219 | 759 | 1709 | 1969 | 2200 | 2393 | 2708 | 2923 |
| genome | 547 | 544 | 532 | 615 | 680 | 797 | 816 | 898 | 1119 | 1280 | 1139 | 1194 | 1369 | 1440 | 1656 | 1744 | 1999 | 2180 | 2288 |
| genomes | 87 | 89 | 59 | 103 | 124 | 163 | 163 | 207 | 236 | 287 | 293 | 232 | 256 | 224 | 313 | 381 | 433 | 404 | 430 |
| genomic | 19 | 28 | 29 | 50 | 77 | 105 | 151 | 276 | 370 | 516 | 644 | 810 | 934 | 1143 | 1468 | 1734 | 1956 | 2198 | 2502 |
| genomics | 0 | 0 | 0 | 0 | 0 | 0 | 0 | 0 | 0 | 0 | 0 | 0 | 41 | 85 | 193 | 231 | 343 | 430 | 396 |
| metagenome | 0 | 0 | 0 | 0 | 0 | 0 | 0 | 0 | 0 | 0 | 0 | 0 | 0 | 0 | 0 | 0 | 0 | 0 | 0 |
| metagenomes | 0 | 0 | 0 | 0 | 0 | 0 | 0 | 0 | 0 | 0 | 0 | 0 | 0 | 0 | 0 | 0 | 0 | 0 | 0 |
| metagenomic | 0 | 0 | 0 | 0 | 0 | 0 | 0 | 0 | 0 | 0 | 0 | 0 | 0 | 0 | 0 | 0 | 0 | 0 | 0 |
| metagenomics | 0 | 0 | 0 | 0 | 0 | 0 | 0 | 0 | 0 | 0 | 0 | 0 | 0 | 0 | 0 | 0 | 0 | 0 | 0 |
| plasmid | 264 | 331 | 362 | 479 | 658 | 744 | 991 | 1257 | 1737 | 1792 | 1870 | 2023 | 2052 | 1994 | 2349 | 2583 | 2735 | 2703 | 2606 |
| plasmids | 73 | 126 | 222 | 165 | 259 | 359 | 385 | 552 | 849 | 768 | 862 | 829 | 714 | 831 | 864 | 853 | 988 | 953 | 860 |
| pseudogene | 0 | 0 | 5 | 2 | 0 | 5 | 21 | 15 | 21 | 53 | 38 | 38 | 61 | 65 | 79 | 70 | 110 | 115 | 110 |
| pseudogenes | 0 | 0 | 0 | 0 | 0 | 5 | 26 | 7 | 16 | 34 | 53 | 32 | 26 | 52 | 63 | 62 | 56 | 66 | 73 |
| **Total DNA-centric** | 18347 | 19258 | 17430 | 18674 | 21785 | 24857 | 26562 | 31041 | 39696 | 44751 | 49033 | 53580 | 61332 | 69643 | 84469 | 93585 | 102039 | 113700 | 123438 |
| microRNA | 0 | 0 | 0 | 0 | 0 | 0 | 0 | 0 | 0 | 0 | 0 | 0 | 0 | 0 | 0 | 0 | 0 | 0 | 0 |
| microRNAs | 0 | 0 | 0 | 0 | 0 | 0 | 0 | 0 | 0 | 0 | 0 | 0 | 0 | 0 | 0 | 0 | 0 | 0 | 0 |
| miRNA | 0 | 0 | 0 | 0 | 0 | 0 | 0 | 0 | 0 | 0 | 0 | 0 | 0 | 0 | 0 | 0 | 0 | 0 | 0 |
| miRNAs | 0 | 0 | 0 | 0 | 0 | 0 | 0 | 0 | 0 | 0 | 0 | 0 | 0 | 0 | 0 | 0 | 0 | 0 | 0 |
| mRNA | 922 | 1214 | 1201 | 1326 | 1301 | 1460 | 1894 | 1682 | 2527 | 2956 | 3400 | 4077 | 5604 | 7281 | 9536 | 11442 | 13375 | 16054 | 17962 |
| mRNAs | 78 | 148 | 133 | 163 | 180 | 218 | 351 | 397 | 603 | 745 | 806 | 871 | 1153 | 1223 | 1453 | 1752 | 2003 | 2287 | 2576 |
| ncRNA | 0 | 0 | 0 | 0 | 0 | 0 | 0 | 0 | 0 | 0 | 0 | 0 | 0 | 0 | 0 | 0 | 0 | 0 | 0 |
| ncRNAs | 0 | 0 | 0 | 0 | 0 | 0 | 0 | 0 | 0 | 0 | 0 | 0 | 0 | 0 | 0 | 0 | 0 | 0 | 0 |
| piRNA | 0 | 0 | 0 | 0 | 0 | 0 | 0 | 0 | 0 | 0 | 0 | 0 | 0 | 0 | 0 | 0 | 0 | 0 | 0 |
| piRNAs | 0 | 0 | 0 | 0 | 0 | 0 | 0 | 0 | 0 | 0 | 0 | 0 | 0 | 0 | 0 | 0 | 0 | 0 | 0 |
| ribozyme | 0 | 0 | 0 | 0 | 0 | 0 | 0 | 0 | 1 | 0 | 1 | 2 | 1 | 1 | 19 | 7 | 27 | 97 | 90 |
| ribozymes | 0 | 0 | 0 | 0 | 0 | 0 | 0 | 0 | 0 | 0 | 0 | 1 | 1 | 0 | 10 | 10 | 14 | 41 | 41 |
| RNA | 5399 | 5480 | 4436 | 4509 | 4576 | 4370 | 4464 | 4553 | 5214 | 5257 | 5299 | 4958 | 5414 | 6048 | 6589 | 7102 | 7486 | 8254 | 9040 |
| RNAs | 172 | 220 | 210 | 256 | 346 | 325 | 438 | 396 | 534 | 566 | 551 | 531 | 636 | 644 | 571 | 634 | 704 | 707 | 812 |
| rRNA | 340 | 462 | 352 | 381 | 419 | 300 | 335 | 453 | 585 | 526 | 471 | 315 | 438 | 499 | 567 | 509 | 555 | 590 | 719 |
| rRNAs | 15 | 39 | 31 | 29 | 31 | 31 | 48 | 42 | 21 | 22 | 16 | 26 | 25 | 43 | 35 | 23 | 31 | 36 | 36 |
| siRNA | 0 | 0 | 0 | 0 | 0 | 0 | 0 | 0 | 0 | 0 | 0 | 0 | 0 | 0 | 0 | 0 | 3 | 0 | 0 |
| siRNAs | 0 | 0 | 0 | 0 | 0 | 0 | 0 | 0 | 0 | 0 | 0 | 0 | 0 | 0 | 0 | 0 | 0 | 0 | 0 |
| snoRNA | 0 | 0 | 0 | 0 | 0 | 0 | 0 | 0 | 0 | 0 | 0 | 0 | 0 | 0 | 0 | 0 | 0 | 9 | 9 |
| snoRNAs | 0 | 0 | 0 | 0 | 0 | 0 | 0 | 0 | 0 | 0 | 0 | 0 | 0 | 0 | 0 | 0 | 2 | 4 | 13 |
| snRNA | 1 | 0 | 3 | 2 | 11 | 4 | 35 | 28 | 64 | 32 | 28 | 69 | 68 | 58 | 84 | 104 | 115 | 147 | 126 |
| snRNAs | 0 | 0 | 7 | 10 | 10 | 0 | 20 | 4 | 33 | 21 | 28 | 28 | 41 | 41 | 29 | 39 | 26 | 38 | 54 |
| tmRNA | 0 | 0 | 0 | 0 | 0 | 0 | 0 | 0 | 0 | 0 | 0 | 0 | 0 | 0 | 0 | 0 | 0 | 0 | 0 |
| tmRNAs | 0 | 0 | 0 | 0 | 0 | 0 | 0 | 0 | 0 | 0 | 0 | 0 | 0 | 0 | 0 | 0 | 0 | 0 | 0 |
| transcript | 13 | 29 | 42 | 54 | 58 | 55 | 88 | 93 | 163 | 202 | 334 | 391 | 604 | 724 | 868 | 1059 | 1113 | 1367 | 1647 |
| transcriptome | 0 | 0 | 0 | 0 | 0 | 0 | 0 | 0 | 0 | 0 | 0 | 0 | 0 | 0 | 0 | 0 | 0 | 0 | 0 |
| transcriptomes | 0 | 0 | 0 | 0 | 0 | 0 | 0 | 0 | 0 | 0 | 0 | 0 | 0 | 0 | 0 | 0 | 0 | 0 | 0 |
| transcriptomic | 0 | 0 | 0 | 0 | 0 | 0 | 0 | 0 | 0 | 0 | 0 | 0 | 0 | 0 | 0 | 0 | 0 | 0 | 0 |
| transcriptomics | 0 | 0 | 0 | 0 | 0 | 0 | 0 | 0 | 0 | 0 | 0 | 0 | 0 | 0 | 0 | 0 | 0 | 0 | 0 |
| transcripts | 80 | 67 | 92 | 151 | 144 | 173 | 240 | 254 | 443 | 587 | 699 | 770 | 1077 | 1355 | 1663 | 1932 | 2277 | 2697 | 3134 |
| tRNA | 684 | 706 | 418 | 555 | 548 | 502 | 492 | 629 | 740 | 677 | 564 | 615 | 697 | 732 | 779 | 785 | 881 | 1031 | 874 |
| tRNAs | 40 | 52 | 50 | 76 | 77 | 132 | 66 | 92 | 98 | 91 | 57 | 63 | 64 | 70 | 98 | 87 | 105 | 89 | 87 |
| **Total RNA-centric** | 7744 | 8417 | 6975 | 7512 | 7701 | 7570 | 8471 | 8623 | 11026 | 11682 | 12254 | 12717 | 15823 | 18719 | 22301 | 25485 | 28717 | 33448 | 37220 |
| antibodies | 2415 | 2395 | 2280 | 2391 | 2579 | 3263 | 4564 | 5850 | 8576 | 10236 | 11337 | 12034 | 12504 | 12926 | 14352 | 14696 | 14640 | 14909 | 15278 |
| antibody | 3771 | 3679 | 3123 | 2992 | 3522 | 4169 | 5137 | 5994 | 8222 | 9288 | 10276 | 11082 | 11816 | 12004 | 12965 | 13444 | 13968 | 13991 | 13898 |
| chaperone | 0 | 1 | 0 | 0 | 0 | 0 | 0 | 0 | 0 | 0 | 0 | 0 | 0 | 2 | 8 | 14 | 26 | 65 | 125 |
| chaperones | 0 | 0 | 0 | 0 | 0 | 0 | 0 | 0 | 0 | 0 | 0 | 0 | 0 | 1 | 3 | 19 | 34 | 32 | 69 |
| enzymatic | 667 | 701 | 506 | 566 | 629 | 707 | 731 | 765 | 963 | 1061 | 1089 | 1079 | 1098 | 1232 | 1293 | 1430 | 1381 | 1457 | 1487 |
| enzyme | 5630 | 6060 | 4830 | 4896 | 5519 | 6107 | 6687 | 7179 | 9544 | 9588 | 9997 | 10208 | 10355 | 10426 | 11474 | 11983 | 12058 | 12400 | 12535 |
| enzymes | 2583 | 2588 | 1974 | 1974 | 2023 | 2387 | 2360 | 2524 | 3038 | 3368 | 3341 | 3280 | 3365 | 3189 | 3742 | 4213 | 3955 | 4112 | 4077 |
| lipoprotein | 292 | 316 | 311 | 345 | 356 | 459 | 670 | 563 | 735 | 813 | 839 | 925 | 1023 | 962 | 1132 | 1138 | 1376 | 1422 | 1356 |
| lipoproteins | 144 | 168 | 146 | 163 | 190 | 247 | 315 | 378 | 423 | 488 | 564 | 469 | 500 | 563 | 635 | 656 | 616 | 685 | 561 |
| peptidase | 33 | 36 | 38 | 50 | 65 | 78 | 77 | 79 | 160 | 174 | 138 | 157 | 206 | 135 | 179 | 219 | 226 | 290 | 221 |
| peptidases | 11 | 26 | 20 | 15 | 32 | 15 | 11 | 22 | 48 | 60 | 52 | 39 | 41 | 28 | 36 | 44 | 59 | 61 | 56 |
| peptide | 513 | 558 | 638 | 591 | 800 | 1139 | 1260 | 1640 | 2185 | 2712 | 3149 | 3636 | 4663 | 5538 | 6775 | 6873 | 8183 | 9295 | 9702 |
| peptides | 317 | 377 | 359 | 378 | 556 | 852 | 778 | 1013 | 1394 | 1682 | 2092 | 2171 | 2340 | 2629 | 3366 | 3806 | 4207 | 4655 | 5044 |
| phosphoprotein | 67 | 47 | 56 | 40 | 86 | 88 | 105 | 116 | 170 | 211 | 248 | 235 | 205 | 297 | 295 | 287 | 361 | 299 | 304 |
| phosphoproteins | 18 | 23 | 29 | 11 | 36 | 50 | 66 | 47 | 122 | 110 | 142 | 150 | 118 | 181 | 176 | 174 | 136 | 181 | 157 |
| polypeptide | 589 | 620 | 584 | 652 | 809 | 1063 | 1249 | 1327 | 1806 | 2014 | 2110 | 2253 | 2407 | 2480 | 2480 | 2634 | 2640 | 2586 | 2534 |
| polypeptides | 419 | 467 | 548 | 583 | 714 | 1014 | 1059 | 1141 | 1442 | 1555 | 1441 | 1448 | 1583 | 1395 | 1518 | 1454 | 1393 | 1279 | 1281 |
| protease | 265 | 362 | 281 | 338 | 443 | 508 | 646 | 708 | 988 | 841 | 1022 | 1198 | 1142 | 1322 | 1453 | 1666 | 1778 | 1923 | 1916 |
| proteases | 123 | 118 | 128 | 123 | 143 | 190 | 214 | 172 | 314 | 317 | 364 | 394 | 425 | 443 | 497 | 507 | 571 | 620 | 623 |
| proteasome | 0 | 0 | 0 | 0 | 0 | 0 | 0 | 0 | 0 | 0 | 0 | 0 | 0 | 2 | 23 | 22 | 56 | 113 | 196 |
| proteasomes | 0 | 0 | 0 | 0 | 0 | 0 | 0 | 0 | 0 | 0 | 0 | 0 | 0 | 12 | 29 | 33 | 33 | 51 | 97 |
| protein | 8292 | 8514 | 7839 | 8165 | 9171 | 10866 | 12838 | 13893 | 18335 | 20750 | 23460 | 26246 | 28623 | 33463 | 39559 | 44019 | 47471 | 53765 | 58028 |
| proteinase | 98 | 117 | 68 | 87 | 171 | 221 | 209 | 277 | 359 | 540 | 466 | 621 | 703 | 802 | 904 | 788 | 1015 | 1003 | 1055 |
| proteinases | 46 | 59 | 32 | 48 | 93 | 100 | 96 | 88 | 128 | 184 | 158 | 212 | 193 | 241 | 266 | 295 | 296 | 301 | 295 |
| proteins | 3738 | 3847 | 3418 | 3664 | 4301 | 4987 | 5607 | 6290 | 7804 | 8873 | 9254 | 10223 | 11185 | 12741 | 15184 | 16744 | 18493 | 20553 | 21693 |
| proteome | 0 | 0 | 0 | 0 | 0 | 0 | 0 | 0 | 0 | 0 | 0 | 0 | 0 | 0 | 0 | 0 | 0 | 0 | 0 |
| proteomes | 0 | 0 | 0 | 0 | 0 | 0 | 0 | 0 | 0 | 0 | 0 | 0 | 0 | 0 | 0 | 0 | 0 | 0 | 0 |
| proteomic | 0 | 0 | 0 | 0 | 0 | 0 | 0 | 0 | 0 | 0 | 0 | 0 | 0 | 0 | 0 | 0 | 0 | 0 | 0 |
| proteomics | 0 | 0 | 0 | 0 | 0 | 0 | 0 | 0 | 0 | 0 | 0 | 0 | 0 | 0 | 0 | 0 | 0 | 0 | 0 |
| secretome | 0 | 0 | 0 | 0 | 0 | 0 | 0 | 0 | 0 | 0 | 0 | 0 | 0 | 0 | 0 | 0 | 0 | 0 | 0 |
| secretomes | 0 | 0 | 0 | 0 | 0 | 0 | 0 | 0 | 0 | 0 | 0 | 0 | 0 | 0 | 0 | 0 | 0 | 0 | 0 |
| secretomic | 0 | 0 | 0 | 0 | 0 | 0 | 0 | 0 | 0 | 0 | 0 | 0 | 0 | 0 | 0 | 0 | 0 | 0 | 0 |
| secretomics | 0 | 0 | 0 | 0 | 0 | 0 | 0 | 0 | 0 | 0 | 0 | 0 | 0 | 0 | 0 | 0 | 0 | 0 | 0 |
| **Total protein-centric** | 30031 | 31079 | 27208 | 28072 | 32238 | 38510 | 44679 | 50066 | 66756 | 74865 | 81539 | 88060 | 94495 | 103014 | 118344 | 127158 | 134972 | 146048 | 152588 |

| | 1994 | 1995 | 1996 | 1997 | 1998 | 1999 | 2000 | 2001 | 2002 | 2003 | 2004 | 2005 | 2006 | 2007 | 2008 | 2009 | 2010 | 2011 |
|---|---|---|---|---|---|---|---|---|---|---|---|---|---|---|---|---|---|---|
| **Publications ("cell")** | 105841 | 108746 | 112407 | 114908 | 119936 | 123355 | 131329 | 135394 | 138484 | 145045 | 153876 | 158447 | 163913 | 168999 | 175871 | 181714 | 191652 | 200718 |
| **Total words** | 26075605 | 27251612 | 28953715 | 29882048 | 31299937 | 32420790 | 34412018 | 35545593 | 36414253 | 38268984 | 40754339 | 41893009 | 43283478 | 44726598 | 47285437 | 48931004 | 52446905 | 55630518 |
| cDNA | 7402 | 7785 | 7562 | 7467 | 7276 | 6889 | 7044 | 6549 | 6165 | 5769 | 5770 | 5037 | 4473 | 3755 | 3182 | 2835 | 2616 | 2423 |
| cDNAs | 1126 | 1165 | 1182 | 1148 | 1102 | 1170 | 1062 | 913 | 844 | 815 | 750 | 602 | 513 | 422 | 428 | 287 | 274 | 270 |
| chromatin | 1609 | 1865 | 1941 | 2237 | 2531 | 2451 | 2896 | 3180 | 3475 | 3704 | 4457 | 4798 | 5333 | 5703 | 5758 | 6435 | 6846 | 7295 |
| chromatins | 12 | 9 | 2 | 3 | 3 | 7 | 1 | 3 | 1 | 2 | 7 | 7 | 14 | 4 | 6 | 7 | 3 | 7 |
| chromosomal | 2708 | 3032 | 3147 | 3178 | 3550 | 3339 | 3708 | 3811 | 3753 | 3870 | 3948 | 4010 | 4053 | 4007 | 3835 | 3636 | 3849 | 3762 |
| chromosome | 10072 | 10847 | 10846 | 10735 | 10958 | 10924 | 11423 | 11514 | 10751 | 10676 | 11033 | 11037 | 10929 | 10322 | 10322 | 9641 | 10158 | 9847 |
| chromosomes | 3111 | 3066 | 3327 | 3193 | 3505 | 3522 | 3845 | 3968 | 3687 | 3883 | 4132 | 3873 | 3761 | 3660 | 3604 | 3680 | 3544 | 3766 |
| DNA | 31209 | 32088 | 33030 | 34318 | 34905 | 36155 | 36985 | 38609 | 37546 | 38118 | 40454 | 41088 | 41493 | 40863 | 41101 | 43007 | 44187 | 46251 |
| DNAs | 448 | 388 | 397 | 363 | 350 | 293 | 289 | 239 | 311 | 271 | 283 | 239 | 238 | 238 | 184 | 234 | 224 | 206 |
| epigenetic | 104 | 111 | 154 | 168 | 221 | 225 | 359 | 440 | 608 | 843 | 1015 | 1462 | 1841 | 2276 | 2688 | 3210 | 4164 | 4692 |
| epigenetics | 2 | 3 | 1 | 1 | 3 | 2 | 13 | 19 | 38 | 46 | 61 | 110 | 138 | 202 | 281 | 285 | 465 | 615 |
| epigenome | 0 | 0 | 0 | 0 | 0 | 0 | 1 | 2 | 3 | 9 | 9 | 27 | 34 | 56 | 66 | 84 | 156 | 180 |
| epigenomes | 0 | 0 | 0 | 0 | 0 | 0 | 0 | 0 | 1 | 2 | 0 | 3 | 1 | 2 | 10 | 14 | 21 | 23 |
| epigenomic | 0 | 0 | 0 | 0 | 0 | 1 | 1 | 0 | 4 | 13 | 6 | 4 | 13 | 19 | 33 | 55 | 96 | 118 |
| epigenomics | 0 | 0 | 0 | 0 | 0 | 0 | 0 | 2 | 9 | 2 | 4 | 9 | 9 | 22 | 25 | 58 | 87 | 104 |
| exome | 0 | 0 | 0 | 0 | 0 | 0 | 0 | 0 | 0 | 0 | 0 | 0 | 0 | 0 | 0 | 2 | 51 | 204 |
| exomes | 0 | 0 | 0 | 0 | 0 | 0 | 0 | 0 | 0 | 0 | 0 | 0 | 0 | 0 | 0 | 0 | 4 | 14 |
| exomic | 0 | 0 | 0 | 0 | 0 | 0 | 0 | 0 | 0 | 0 | 0 | 0 | 0 | 0 | 0 | 0 | 1 | 3 |
| exomics | 0 | 0 | 0 | 0 | 0 | 0 | 0 | 0 | 0 | 0 | 0 | 0 | 0 | 0 | 0 | 0 | 0 | 0 |
| gene | 44437 | 47496 | 50484 | 53787 | 55622 | 57963 | 60935 | 62270 | 63347 | 65451 | 69127 | 69163 | 69959 | 69654 | 70312 | 70197 | 72499 | 74736 |
| genes | 14914 | 16109 | 17185 | 18644 | 20194 | 20988 | 23739 | 26274 | 28614 | 31613 | 35184 | 37014 | 38968 | 40165 | 40784 | 41359 | 43616 | 45367 |
| genetic | 5603 | 6187 | 6645 | 7211 | 8185 | 8871 | 10071 | 10794 | 11154 | 11915 | 12899 | 14062 | 14434 | 15283 | 16015 | 17220 | 18472 | 20384 |
| genetics | 2993 | 3271 | 3423 | 3700 | 4193 | 4471 | 5193 | 5450 | 5364 | 5725 | 6107 | 6363 | 6469 | 6600 | 7062 | 7078 | 7250 | 7846 |
| genome | 2390 | 2741 | 3022 | 3406 | 3516 | 3825 | 4615 | 5542 | 6126 | 6893 | 8036 | 8608 | 9252 | 9615 | 10284 | 11139 | 12285 | 13622 |
| genomes | 441 | 505 | 504 | 619 | 618 | 712 | 755 | 1017 | 1054 | 1282 | 1430 | 1690 | 1674 | 1804 | 1878 | 1959 | 2016 | 2168 |
| genomic | 2745 | 2988 | 3146 | 3391 | 3691 | 3765 | 4172 | 4652 | 4643 | 4872 | 5360 | 5594 | 5897 | 5820 | 6413 | 6458 | 6854 | 7141 |
| genomics | 468 | 459 | 460 | 455 | 445 | 481 | 606 | 835 | 1029 | 1101 | 1536 | 1665 | 1854 | 2016 | 2149 | 2185 | 2374 | 2385 |
| metagenome | 0 | 0 | 0 | 0 | 0 | 0 | 1 | 0 | 1 | 1 | 6 | 7 | 14 | 18 | 13 | 30 | 15 | 41 |
| metagenomes | 0 | 0 | 0 | 0 | 0 | 0 | 0 | 0 | 0 | 0 | 0 | 2 | 2 | 2 | 5 | 6 | 7 | 18 |
| metagenomic | 0 | 0 | 0 | 0 | 0 | 0 | 2 | 0 | 0 | 3 | 0 | 17 | 19 | 41 | 28 | 42 | 77 | 120 |
| metagenomics | 0 | 0 | 0 | 0 | 0 | 0 | 0 | 0 | 0 | 0 | 2 | 8 | 3 | 13 | 25 | 26 | 22 | 31 |
| plasmid | 2603 | 2638 | 2787 | 2711 | 2931 | 3197 | 3163 | 3282 | 3170 | 3448 | 3554 | 3888 | 3859 | 3684 | 3519 | 3427 | 3388 | 3680 |
| plasmids | 934 | 874 | 887 | 894 | 918 | 938 | 975 | 930 | 913 | 1071 | 1023 | 1016 | 1039 | 1070 | 1021 | 1037 | 1023 | 1125 |
| pseudogene | 129 | 135 | 110 | 129 | 129 | 135 | 121 | 121 | 118 | 68 | 75 | 116 | 75 | 70 | 84 | 55 | 58 | 58 |
| pseudogenes | 47 | 70 | 87 | 54 | 80 | 89 | 81 | 128 | 148 | 93 | 112 | 124 | 115 | 100 | 123 | 54 | 89 | 78 |
| **Total DNA-centric** | 135507 | 143832 | 150329 | 157812 | 164926 | 170413 | 182056 | 190544 | 192877 | 201559 | 216380 | 221643 | 226476 | 227506 | 231238 | 235742 | 246791 | 258580 |
| microRNA | 0 | 0 | 0 | 0 | 0 | 0 | 0 | 0 | 6 | 36 | 101 | 181 | 274 | 507 | 928 | 1374 | 2169 | 2644 |
| microRNAs | 0 | 0 | 0 | 0 | 0 | 0 | 0 | 3 | 18 | 51 | 116 | 231 | 356 | 613 | 1060 | 1463 | 2083 | 2480 |
| miRNA | 0 | 0 | 0 | 0 | 0 | 0 | 0 | 0 | 20 | 70 | 171 | 346 | 639 | 1019 | 1500 | 2369 | 3316 | 4265 |
| miRNAs | 0 | 0 | 0 | 0 | 0 | 0 | 0 | 0 | 41 | 100 | 197 | 384 | 778 | 1192 | 1691 | 2744 | 3963 | 4838 |
| mRNA | 20957 | 22524 | 24222 | 24168 | 24923 | 24357 | 24892 | 24975 | 23816 | 24277 | 24813 | 25106 | 25031 | 25163 | 25041 | 24700 | 24948 | 25303 |
| mRNAs | 2915 | 2703 | 2926 | 2637 | 2597 | 2610 | 2696 | 2647 | 2417 | 2482 | 2367 | 2485 | 2259 | 2327 | 2583 | 2505 | 2400 | 2584 |
| ncRNA | 0 | 0 | 0 | 0 | 0 | 1 | 6 | 1 | 0 | 7 | 0 | 16 | 38 | 49 | 54 | 82 | 86 | 107 |
| ncRNAs | 0 | 0 | 0 | 0 | 0 | 0 | 0 | 4 | 4 | 8 | 2 | 33 | 63 | 66 | 69 | 142 | 165 | 173 |
| piRNA | 0 | 0 | 0 | 0 | 0 | 0 | 0 | 0 | 0 | 1 | 0 | 0 | 4 | 30 | 48 | 90 | 85 | 156 |
| piRNAs | 0 | 0 | 0 | 0 | 0 | 0 | 0 | 0 | 0 | 0 | 0 | 0 | 20 | 62 | 76 | 100 | 90 | 124 |
| ribozyme | 253 | 206 | 267 | 294 | 370 | 283 | 457 | 357 | 380 | 328 | 395 | 301 | 222 | 193 | 123 | 149 | 110 | 98 |
| ribozymes | 125 | 128 | 158 | 187 | 176 | 192 | 219 | 181 | 254 | 168 | 201 | 156 | 90 | 92 | 42 | 60 | 56 | 62 |
| RNA | 9376 | 10260 | 10449 | 10833 | 10713 | 10919 | 11124 | 10963 | 11055 | 11840 | 13765 | 15068 | 15796 | 16473 | 17390 | 17672 | 18385 | 19666 |
| RNAs | 823 | 927 | 802 | 954 | 835 | 775 | 879 | 859 | 919 | 1148 | 1313 | 1550 | 1811 | 1861 | 2065 | 2329 | 2418 | 2686 |
| rRNA | 766 | 949 | 821 | 1029 | 1036 | 1084 | 1127 | 1015 | 976 | 1073 | 1335 | 1380 | 1419 | 1626 | 1639 | 1472 | 1602 | 1727 |
| rRNAs | 60 | 49 | 47 | 61 | 51 | 48 | 70 | 57 | 43 | 50 | 48 | 47 | 62 | 44 | 54 | 60 | 62 | 77 |
| siRNA | 0 | 0 | 1 | 0 | 0 | 0 | 1 | 12 | 188 | 733 | 1640 | 2681 | 3822 | 4642 | 5340 | 5930 | 6777 | 7398 |
| siRNAs | 0 | 0 | 0 | 0 | 0 | 0 | 0 | 25 | 129 | 349 | 536 | 737 | 782 | 867 | 867 | 916 | 915 | 1026 |
| snoRNA | 12 | 27 | 44 | 60 | 103 | 57 | 54 | 53 | 73 | 37 | 62 | 27 | 33 | 28 | 61 | 71 | 62 | 47 |
| snoRNAs | 10 | 47 | 34 | 52 | 80 | 69 | 52 | 50 | 68 | 27 | 20 | 22 | 34 | 41 | 41 | 71 | 64 | 61 |
| snRNA | 105 | 149 | 133 | 99 | 159 | 149 | 86 | 87 | 130 | 111 | 121 | 129 | 118 | 88 | 104 | 108 | 87 | 87 |
| snRNAs | 38 | 43 | 49 | 36 | 29 | 48 | 26 | 22 | 47 | 33 | 43 | 44 | 56 | 38 | 60 | 50 | 52 | 44 |
| tmRNA | 0 | 0 | 0 | 7 | 7 | 11 | 33 | 62 | 53 | 49 | 54 | 70 | 69 | 94 | 45 | 105 | 89 | 79 |
| tmRNAs | 0 | 0 | 0 | 0 | 0 | 0 | 0 | 0 | 0 | 0 | 2 | 1 | 0 | 2 | 0 | 3 | 0 | 2 |
| transcript | 1926 | 1929 | 2095 | 2165 | 2342 | 2322 | 2169 | 2173 | 2195 | 2382 | 2384 | 2349 | 2390 | 2662 | 2476 | 2520 | 2714 | 2863 |
| transcriptome | 0 | 0 | 0 | 2 | 1 | 3 | 18 | 72 | 131 | 257 | 349 | 479 | 628 | 706 | 858 | 926 | 1319 | 1525 |
| transcriptomes | 0 | 0 | 0 | 0 | 0 | 3 | 5 | 6 | 26 | 27 | 33 | 65 | 93 | 98 | 105 | 105 | 184 | 220 |
| transcriptomic | 0 | 0 | 0 | 0 | 0 | 0 | 1 | 8 | 11 | 18 | 33 | 42 | 103 | 127 | 216 | 282 | 393 | 448 |
| transcriptomics | 0 | 0 | 0 | 0 | 0 | 0 | 0 | 3 | 8 | 7 | 24 | 32 | 47 | 62 | 73 | 91 | 144 | 186 |
| transcripts | 3444 | 3608 | 3706 | 3764 | 3670 | 3556 | 3458 | 3357 | 3378 | 3285 | 3489 | 3574 | 3617 | 3669 | 3735 | 3610 | 3753 | 3907 |
| tRNA | 1178 | 1029 | 1111 | 1043 | 964 | 871 | 1111 | 925 | 1127 | 1024 | 1154 | 1136 | 1072 | 1195 | 1253 | 1112 | 1202 | 1131 |
| tRNAs | 114 | 137 | 149 | 125 | 184 | 118 | 158 | 158 | 144 | 158 | 182 | 175 | 228 | 218 | 211 | 207 | 213 | 223 |
| **Total RNA-centric** | 42102 | 44715 | 47014 | 47516 | 48240 | 47476 | 48647 | 48078 | 47657 | 50136 | 54950 | 58847 | 61954 | 65854 | 69908 | 73418 | 79906 | 86237 |
| antibodies | 15082 | 14653 | 14173 | 13702 | 13347 | 12647 | 12336 | 12039 | 11368 | 11169 | 11823 | 11162 | 10905 | 10676 | 10893 | 10938 | 10955 | 11391 |
| antibody | 14403 | 14248 | 13892 | 13264 | 13327 | 12714 | 12817 | 12832 | 11683 | 11734 | 12926 | 11971 | 11806 | 12175 | 12173 | 12249 | 12719 | 13288 |
| chaperone | 231 | 249 | 342 | 370 | 489 | 574 | 628 | 766 | 728 | 778 | 1056 | 1052 | 1124 | 1200 | 1283 | 1233 | 1490 | 1557 |
| chaperones | 143 | 135 | 238 | 247 | 284 | 299 | 395 | 412 | 450 | 473 | 569 | 563 | 592 | 692 | 628 | 701 | 915 | 784 |
| enzymatic | 1630 | 1716 | 1676 | 1742 | 1901 | 1799 | 1960 | 1968 | 2014 | 2092 | 2288 | 2416 | 2425 | 2449 | 2628 | 2634 | 2778 | 2899 |
| enzyme | 12441 | 12743 | 13182 | 12811 | 12738 | 12818 | 12996 | 12395 | 12770 | 12658 | 12869 | 12593 | 12271 | 12186 | 12238 | 11943 | 12375 | 12805 |
| enzymes | 4389 | 4458 | 4562 | 4548 | 4902 | 5146 | 5384 | 5696 | 5685 | 6060 | 6332 | 6441 | 6772 | 6718 | 7131 | 7111 | 7782 | 8097 |
| lipoprotein | 1514 | 1596 | 1578 | 1609 | 1649 | 1638 | 1740 | 1622 | 1663 | 1617 | 1811 | 1612 | 1795 | 1710 | 1692 | 1778 | 1836 | 1965 |
| lipoproteins | 737 | 728 | 601 | 720 | 724 | 807 | 679 | 660 | 686 | 578 | 650 | 562 | 579 | 546 | 542 | 555 | 561 | 568 |
| peptidase | 303 | 247 | 286 | 266 | 336 | 305 | 307 | 239 | 297 | 312 | 357 | 322 | 414 | 412 | 443 | 449 | 367 | 437 |
| peptidases | 70 | 50 | 57 | 59 | 54 | 53 | 62 | 58 | 77 | 90 | 96 | 90 | 108 | 121 | 122 | 120 | 88 | 85 |
| peptide | 10522 | 11020 | 11469 | 11126 | 11558 | 11551 | 11972 | 11795 | 12155 | 12332 | 14096 | 13039 | 12507 | 12979 | 12850 | 13447 | 13865 | 14592 |
| peptides | 5803 | 5775 | 5907 | 5978 | 5541 | 5641 | 5894 | 5891 | 6242 | 6355 | 7250 | 6824 | 6580 | 6721 | 6941 | 6878 | 7622 | 7975 |
| phosphoprotein | 358 | 356 | 316 | 355 | 283 | 314 | 282 | 274 | 214 | 267 | 279 | 277 | 263 | 338 | 318 | 308 | 340 | 347 |
| phosphoproteins | 131 | 149 | 117 | 124 | 108 | 84 | 72 | 65 | 69 | 75 | 103 | 91 | 101 | 126 | 122 | 112 | 110 | 168 |
| polypeptide | 2715 | 2445 | 2325 | 2174 | 2050 | 1871 | 1735 | 1574 | 1509 | 1364 | 1422 | 1223 | 1235 | 1162 | 1127 | 1159 | 1187 | 1214 |
| polypeptides | 1268 | 1101 | 1026 | 1034 | 829 | 787 | 778 | 769 | 676 | 619 | 604 | 613 | 512 | 473 | 456 | 400 | 448 | 401 |
| protease | 2066 | 2299 | 2686 | 2842 | 3070 | 3112 | 3208 | 3017 | 2998 | 3194 | 3241 | 3173 | 3129 | 3379 | 3230 | 3194 | 3453 | 3492 |
| proteases | 662 | 761 | 1071 | 1285 | 1077 | 1209 | 1146 | 1114 | 1250 | 1202 | 1470 | 1295 | 1320 | 1332 | 1367 | 1350 | 1375 | 1495 |
| proteasome | 219 | 338 | 497 | 711 | 942 | 1277 | 1590 | 1663 | 1797 | 2181 | 2426 | 2720 | 2698 | 2917 | 3025 | 2837 | 3204 | 2920 |
| proteasomes | 95 | 117 | 132 | 151 | 154 | 171 | 226 | 186 | 205 | 182 | 181 | 232 | 248 | 196 | 185 | 171 | 172 | 186 |
| protein | 64140 | 67892 | 71045 | 74821 | 77453 | 82447 | 84963 | 86970 | 87486 | 92597 | 96873 | 98918 | 100429 | 101556 | 102659 | 103809 | 107527 | 112814 |
| proteinase | 956 | 1111 | 992 | 988 | 900 | 723 | 874 | 739 | 719 | 683 | 677 | 647 | 629 | 597 | 605 | 553 | 480 | 474 |
| proteinases | 326 | 383 | 335 | 385 | 403 | 310 | 339 | 289 | 342 | 302 | 313 | 264 | 241 | 238 | 244 | 169 | 152 | 223 |
| proteins | 23935 | 25856 | 27069 | 27983 | 29894 | 31557 | 33459 | 34618 | 35329 | 37565 | 41529 | 42713 | 43854 | 43892 | 46216 | 47188 | 48863 | 51302 |
| proteome | 0 | 0 | 4 | 46 | 38 | 61 | 141 | 234 | 383 | 544 | 766 | 1050 | 1235 | 1373 | 1471 | 1741 | 1886 | 1979 |
| proteomes | 0 | 0 | 0 | 1 | 2 | 10 | 23 | 43 | 64 | 60 | 127 | 128 | 172 | 208 | 198 | 216 | 218 | 264 |
| proteomic | 0 | 0 | 0 | 5 | 6 | 25 | 77 | 204 | 388 | 538 | 1013 | 1140 | 1578 | 1517 | 1788 | 2009 | 2217 | 2381 |
| proteomics | 0 | 0 | 0 | 0 | 3 | 33 | 142 | 298 | 642 | 795 | 1189 | 1422 | 1730 | 1779 | 2013 | 2164 | 2435 | 2348 |
| secretome | 0 | 0 | 0 | 0 | 0 | 0 | 3 | 3 | 3 | 4 | 16 | 22 | 50 | 69 | 129 | 140 | 199 | 224 |
| secretomes | 0 | 0 | 0 | 0 | 0 | 0 | 0 | 0 | 0 | 0 | 5 | 6 | 12 | 8 | 18 | 18 | 33 | 29 |
| secretomic | 0 | 0 | 0 | 0 | 0 | 0 | 0 | 0 | 0 | 0 | 0 | 0 | 0 | 0 | 0 | 2 | 3 | 6 |
| secretomics | 0 | 0 | 0 | 0 | 0 | 0 | 0 | 0 | 0 | 0 | 2 | 0 | 2 | 1 | 1 | 0 | 6 | 5 |
| **Total protein-centric** | 164139 | 170426 | 175578 | 179347 | 184062 | 189983 | 196228 | 198433 | 199892 | 208420 | 224359 | 224581 | 227316 | 229746 | 234736 | 237576 | 247661 | 258715 |

**Figure S1**

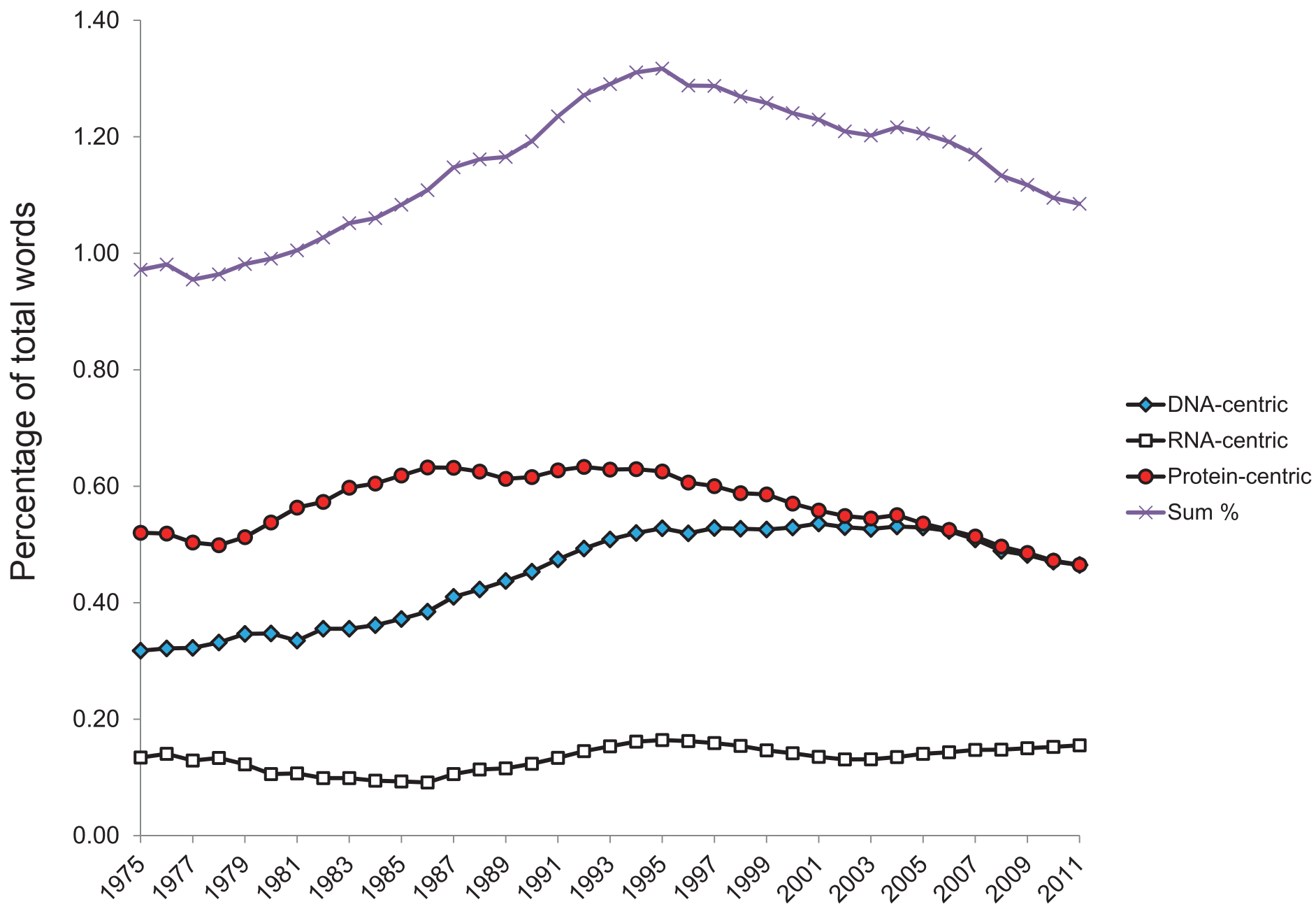
1.40
1.20
1.00
0.80
0.60
0.40
0.20
0.00
Percentage of total words
1975
1977
1979
1981
1983
1985
1987
1989
1991
1993
1995
1997
1999
2001
2003
2005
2007
2009
2011
DNA-centric
RNA-centric
Protein-centric
Sum %

**Figure S2**

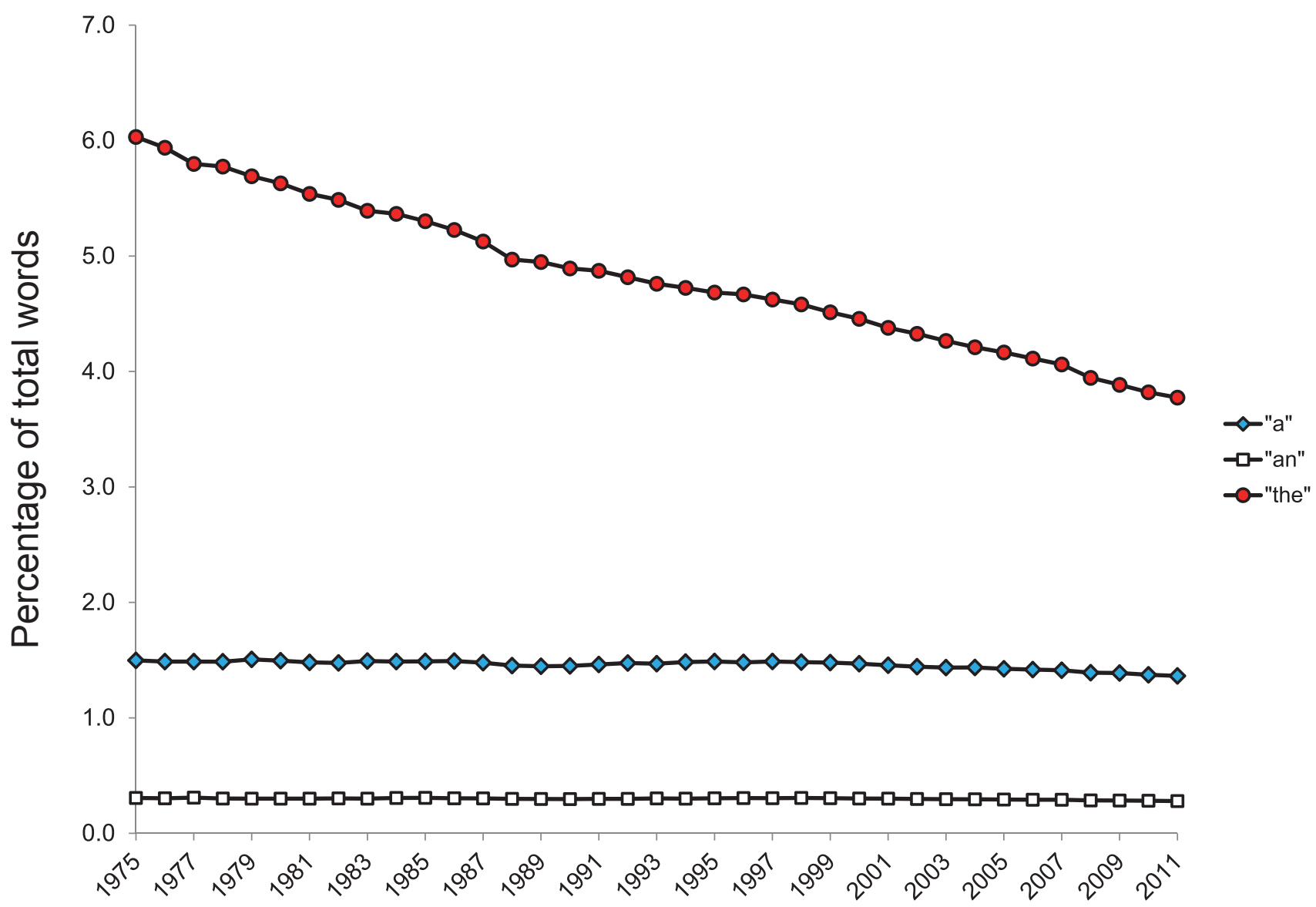

7.0
6.0
5.0
4.0
3.0
2.0
1.0
0.0
Percentage of total words
1975
1977
1979
1981
1983
1985
1987
1989
1991
1993
1995
1997
1999
2001
2003
2005
2007
2009
2011
"a"
"an"
"the"

**Figure S3**

**A**

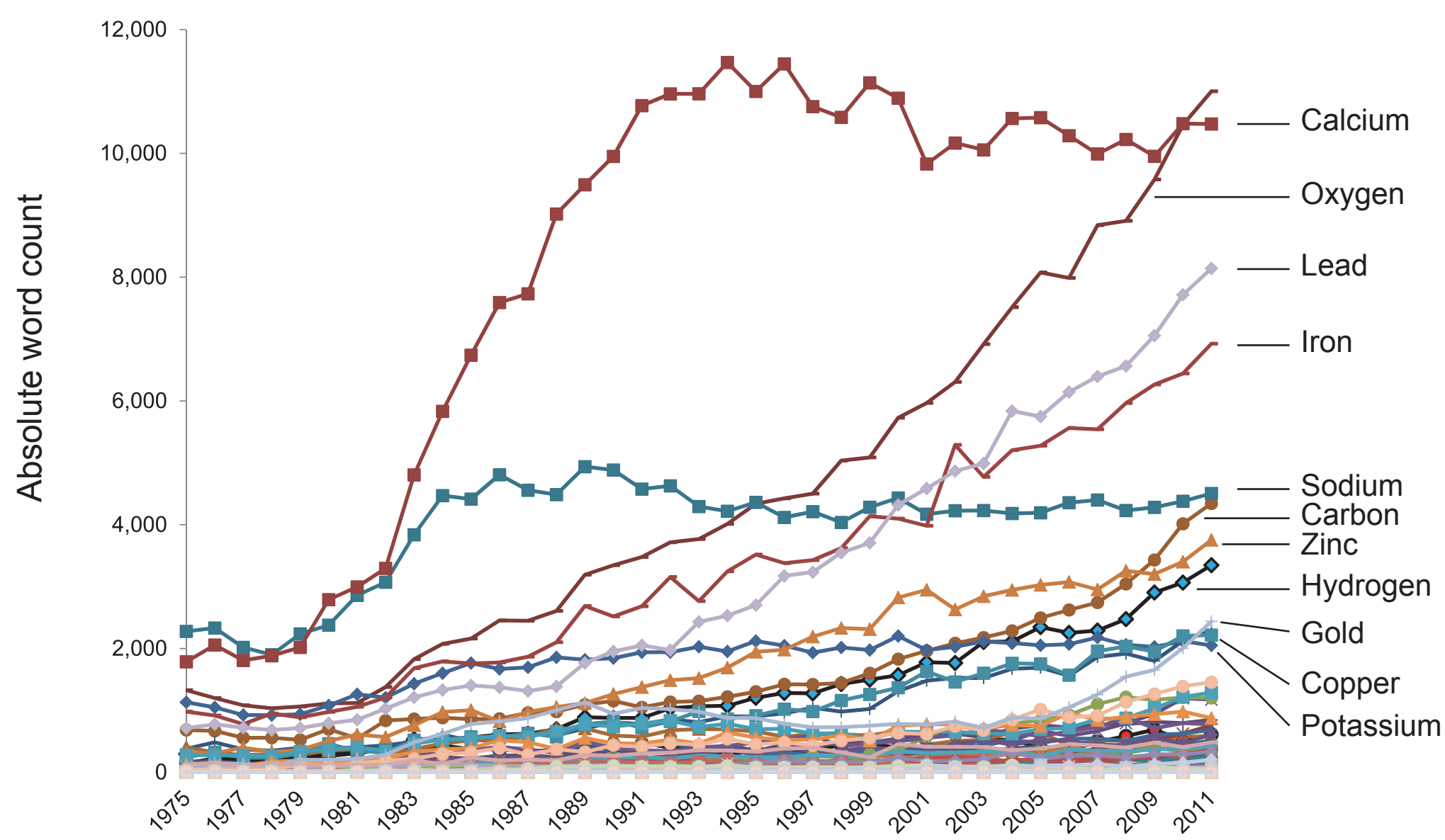
Absolute word count
12,000
10,000
8,000
6,000
4,000
2,000
0
1975
1977
1979
1981
1983
1985
1987
1989
1991
1993
1995
1997
1999
2001
2003
2005
2007
2009
2011
Calcium
Oxygen
Lead
Iron
Sodium
Carbon
Zinc
Hydrogen
Gold
Copper
Potassium

**B**

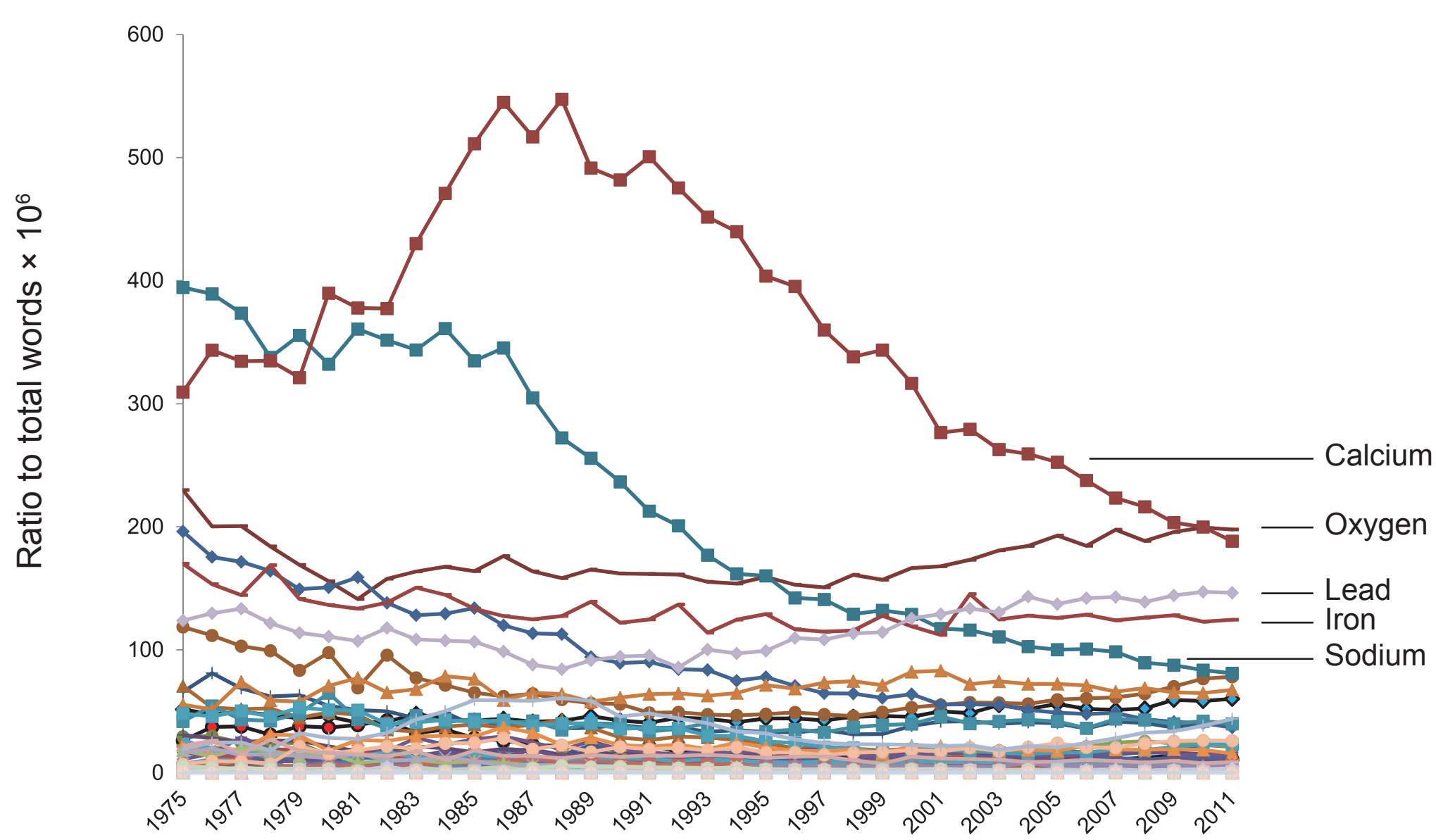
Ratio to total words × 10⁶
600
500
400
300
200
100
0
1975
1977
1979
1981
1983
1985
1987
1989
1991
1993
1995
1997
1999
2001
2003
2005
2007
2009
2011
Calcium
Oxygen
Lead
Iron
Sodium